\documentclass[11pt,a4paper]{article}

\usepackage{fullpage}

\usepackage[english]{babel}

\usepackage{amsmath}
\usepackage{amsfonts}
\usepackage{amssymb}

\def \be{\begin{equation}}
\def \ee{\end{equation}}

\def \um {\frac{1}{2}}

\author{V. Aldaya$^{1}$, M. Calixto$^{1,2}$, J. Guerrero$^{1,3}$ and F. F. L\'opez-Ruiz$^{1}$}

\title{Group-quantization of non-linear sigma models:\\ particle on $\mathbb S^2$ revisited}

\date{\begin{center}
\begin{small}$^1$ Instituto de Astrof\'\i sica de Andaluc\'\i a, CSIC,             \end{small}\\
\begin{small}
   Apartado Postal 3004, 18080 Granada, Spain
\end{small}\\
\begin{small}$^2$ Departamento de Matem\'atica Aplicada y Estad\'\i stica,\\
Universidad Polit\'ecnica de Cartagena, \end{small}\\
\begin{small}Paseo Alfonso XIII 56, 30203 Cartagena,
Spain\end{small}\\
\begin{small}$^3$Departamento de Matem\'atica Aplicada, Universidad de Murcia, \end{small}\\
\begin{small}Campus de Espinardo, 30100 Murcia, Spain.\end{small}\\
\begin{small}valdaya@iaa.es \ manuel.calixto@upct.es \ juguerre@um.es\ flopez@iaa.es\end{small}\\                                                                                                 \end{center}
}
\begin{document}

\maketitle
%


\begin{abstract}
We present the quantum mechanics of ``partial-trace'' non-linear sigma models, on the grounds of a fully symmetry-based procedure. After the general scheme is sketched, the particular example of a particle on the two-sphere is explicitly
developed. As a remarkable feature, no explicit constraint treatment is
required nor ordering ambiguities do appear. Moreover, the energy
spectrum is recovered without extra terms in the curvature of the sphere.
\end{abstract}
 
11.30.-j, 03.65.-w, 05.45.-a.

\section{Introduction}

The quantization of simple second-class constrained mechanical systems has usually been 
accomplished by adopting as canonical commutation relations those given by the so-called 
Dirac algorithm. This method, although successful in many situations, can potentially lead 
to wrong results, and additional strategies or physical considerations are needed. This, 
which is not a serious problem when the system under consideration contains ``known 
physics'', as in the case of the free particle constrained to move on a sphere surface 
\cite{properdirac, circle, stuecksphere}, could become dramatic if we want to unfold new physical features.

In general, canonical quantization makes extensive use of Hamiltonian formulation of 
classical mechanics, trying to put the classical theory in a form in which the quantization 
would be straightforward and/or as unambiguous as possible. However, it should be 
recalled once again that it is the quantum system what contains the physical entity,
the classical theory being a mere approximation. 

This paper is devoted to show how, in the case of the partial-trace non-linear sigma model (NLSM) at least, a proper quantum theory can be obtained straightforwardly if one leaves aside 
the idea that the classical theory must lead to the quantum one through a process of  
\textit{quantization}, which is inherently ambiguous. Here we suggest that the 
\textit{quantized system} should be \textit{directly} obtained through a deeper knowledge 
of the corresponding dynamical symmetry transformations, i.e. those that are able to span the whole space of physical states 
(and/or parameterize the classical solution manifold by means of Noether invariants). We aim at obtaining the quantum theory not from an appropriate treatment of canonical classical 
quantities, but from a clear algorithm based in the complete symmetry of the system, 
which naturally selects the basic quantum operators. 
This method, named Group Approach to Quantization (GAQ) \cite{23}, has proved to be 
successful for many physical systems (see, for instance \cite{oscilrelativ} and \cite{revisited}).

After introducing the general setting of the problem we shall be explicitly concerned with the example of $SU(2)$.

\section{Sigma Model-type systems}
\label{4}

Generally speaking, the non-linear sigma model (NLSM) consists of a set of coupled 
scalar fields 
$\epsilon^i(x^\mu), i=1,\dots,d,$ in a $D$-dimensional spacetime 
$M, \mu=0,1,2,\dots,D-1$, with Lagrangian (density)
\be
\mathcal L_\Sigma=\frac{1}{2}\kappa 
g_{ij}(\epsilon)\partial^\mu\epsilon^i\partial_\mu\epsilon^j, 
\label{nlsmaction}\ee 
where $\partial^\mu=\eta^{\mu\nu}\partial_\nu, 
\partial_\nu=\partial/\partial x^\nu$, $\eta$ is the spacetime  
metric and $\kappa$ a constant. The field theory 
(\ref{nlsmaction}) is called the NLSM with metric $g_{ij}(\epsilon)$ 
(usually a positive-definite field-dependent matrix). The fields 
$\epsilon^i$ themselves can also be considered as the coordinates of an 
internal (pseudo-)Riemannian manifold $\Sigma$ with metric $g_{ij}$. 

In this letter we shall restrict ourselves to the quantum mechanical case $D=1$, so that 
the fields $\epsilon^i(x^\mu)$ are just curves $\epsilon^i(t)$ on $\Sigma$, which we shall take as a 
(semisimple, linear) Lie group manifold $G$, or a given coset $G/G_\lambda$ (see later). Let us denote by $g=e^{\epsilon^iT_i}$ 
an element of $G$, where $T_i, i=1,\dots,{\rm dim}(G)$, stands for the Lie algebra generators, with commutation relations $[T_i,T_j]=C^{\phantom{ij}k}_{ij} T_k$.  For the sake of simplicity, we shall choose $T_i$ in the adjoint representation, whose matrix elements are $(T_i)^k_j=C^{\phantom{ij}k}_{ij}$. Then the Killing form, used to raise and lower indices, is given by $K_{ij}\equiv C^{\phantom{il}k}_{il} C^{\phantom{jk}l}_{jk}={\rm Tr}_G(T_iT_j)$. With this notation, the NLSM Lagrangian acquires the simple algebraic form:
\be
 \mathcal{L}_{G}=\um \kappa{\rm Tr}_{G}(\theta{\theta})\,,
\label{trazatoa}
\ee
where $\theta\equiv g^{-1}\dot g$ and here $\kappa$ is intended to have the dimensions of an inertia moment.

It is well-known the difficulty found by canonical quantization in dealing 
with non-linear systems. Even symmetry-based quantization techniques face the 
impossibility of parameterizing the solution manifold by a finite dimensional Lie group. 
This is essentially because $g^{-1} \dot g$ is not a total derivative, except for Abelian
groups. However, this obstruction disappears when the manifold $\Sigma$ is considered to be 
a coset $G/G_\lambda$ of $G$, $G_\lambda$ being the isotropy subgroup of a given Lie algebra element $\lambda$ under the adjoint action $\lambda \to g \lambda g^{-1}$. To be precise $\lambda$ should have been defined as an element of the dual of the Lie algebra, which is equivalent to the Lie algebra since $G$ is semisimple. In this sense $\lambda$, which can be seen as an ordinary vector, may be endowed with length dimensions so that $\kappa$ would appear as a mass $m$ rather than an inertia moment.

In this case, the (total-trace) NLSM Lagrangian (\ref{trazatoa}) takes 
the (partial-trace) form:      
\be
 \mathcal{L}_{G/G_\lambda}=\um m {\rm Tr}_{G/G_\lambda}(\theta \theta)
 \equiv\um m {\rm Tr}_{G}(\theta_\lambda \theta_\lambda), \label{trazapar}
\ee 
 where we have defined 
\be
\theta_\lambda\equiv[\theta,\lambda]\,.
\ee

It can be realized that, defining
\[
S \equiv g\lambda g^{-1},\,\,g\in G\,,
\]
we have an alternative way of writing (\ref{trazapar}) as 
\be
\mathcal{L}_{G/G_\lambda}=\um m {\rm Tr}_{G}(\dot S \dot S)=\um m K_{ij}\dot S^i\dot S^j, \label{nesfera}.
\ee 
where $S^i \equiv {\rm Tr}_{G}(T^i S)$. 
Note that this Lagrangian is singular due to the existence of constraints like, for example, ${\rm Tr}_{G}(S^2)=
{\rm Tr}_{G}(\lambda^2)\equiv r^2$. We shall not deal with constraints 
at this stage. They will be naturally addressed inside our quantization procedure below.

We can try to find two sets of generators mimicking the basic symmetry of the Galilean
particle (i.e., translations and boosts). For the first set, we choose the generators of the group itself: 
\be
   X_i = C^{\phantom{ij}k}_{ij} S^j \frac{\partial}{\partial S^k} + C^{\phantom{ij}k}_{ij} \dot S^j
    \frac{\partial}{\partial \dot S^k}\,.
   \label{xsinpunto}
\ee
The Lagrangian (\ref{nesfera}) is \textit{strictly} invariant under the action of these generators, i.e.:

\[
   L_{X_i}\mathcal{L}_{G/G_\lambda} = 0\,,  \qquad  \forall i=1,\dots,{\rm dim}(G)\ ,
\]
where  $L_{X}$ stands for the Lie derivative with respect to a generator $X$. In this computation, the fact that the product $K_{lm} C^{\phantom{ij}l}_{ij}\equiv C_{ijm}$ is fully antisymmetric has been used.

As far as the second set of symmetries is concerned (those playing the role of boosts), we propose the following one:

\be
  X'_i = \frac{\partial}{\partial \dot S^i}\,.
  \label{xpunto}
\ee

These generators leave the Lagrangian \textit{semi}-invariant in the sense that they give a total derivative (thus leaving the action strictly invariant), much in the same way the generators of boosts do in the free Galilean particle, that is,  
\be
   L_{X'_i}\mathcal L_{G/G_\lambda} = m \dot{S}_i , \qquad  \forall i=1,\dots,{\rm dim}(G) \ .
   \label{semiinv}
\ee

The generators (\ref{xsinpunto}) and (\ref{xpunto}) close a finite-dimensional Lie algebra with commutation relations
\be
\left[X_i, X_j\right]= - C_{ij}^{\phantom{ij}k} X_{k},\;\; 
\left[X_i, X'_j\right] = - C_{ij}^{\phantom{ij}k}  X'_k,\;\; 
\left[X'_i, X'_j\right] = 0.
\label{g1}
\ee
The corresponding symmetry group is the (co-)tangent group of $G$ and will be denoted by $G^{(1)}$ (see \cite{jetgaugegroup} as regards gauge theory).

We shall assume this algebra (in fact, a central extension of it) as the basic symmetry of the quantum  particle constrained to move on the manifold $G/G_\lambda$.
In fact, the semi-invariance (\ref{semiinv}) of the Lagrangian suggests the presence of a central extension $\tilde G^{(1)}$ of the group $G^{(1)}$, as happens in the quantum Galilean particle \cite{centralext}. At the Lie algebra level, this central extension only affects the second commutator in (\ref{g1}), which now reads

\be
\left[\tilde X_i, \tilde X'_j\right] = - C_{ij}^{\phantom{ij}k} \tilde X'_k 
- C_{ij}^{\phantom{ij}k} \lambda_k \frac{m}{\hbar} \Xi
\label{gtilde1}
\ee
where $\Xi$ denotes the central generator.

This centrally extended group $\tilde G^{(1)}$ is the group of \textit{strict} invariance of the system, and contains the necessary information to obtain the quantum theory. Although this central extension is trivial from a strict mathematical point of view, in the sense that a redefinition of a generator eliminates the central generator from the r.h.s. of the Lie algebra commutators, physically it behaves as a non-trivial one, since under an In\"on\"u-Wigner contraction (limit process) leads to a non-trivial extension of the contracted group \cite{saletan} (see also \cite{refinement} and references therein).

It should be remarked that the commutation relations providing the central term on the right hand side generalize those of the Heisemberg-Weyl algebra, where $\tilde X_i$ play the role of (non-Abelian) ``translation'' generators, and $\tilde X'_j$ the role of ``boost'' generators, although restricted to the coset space  $G/G_\lambda$. This point will be further clarified in the example of the $SU(2)$ group.

\section{Quantum theory}
\label{3}

The leading idea is to obtain an irreducible and unitary representation (unirep) of the basic group suggested in the previous section, i.e. the (centrally extended) $\tilde G^{(1)}$ . The central extension will select a specific representation, associated with a given coadjoint orbit.

This can be performed in any desired way. Here we shall choose a specific algorithm, GAQ, developed by some of the authors.

\subsection{Group Approach to Quantization}

The basic idea of GAQ consists in taking advantage of having {\it two} mutually {\it commuting} 
copies of the Lie algebra $\tilde{\cal G}$ of a group $\tilde{G}$ ($G$ centrally extendended by $U(1)$) of {\it strict symmetry} of a given physical system, that is,

\[{\cal X}^L(\tilde{G})\approx\tilde{\cal G}\approx{\cal 
X}^R(\tilde{G})\]

\noindent  in such a way that one copy, let us say 
${\cal X}^R(\tilde{G})$, plays the role of {\it pre-Quantum 
Operators} acting (by usual derivation) on complex (wave) functions on 
$\tilde{G}$, whereas the other,
${\cal X}^L(\tilde{G})$, is used to {\it reduce} the 
representation in a manner {\it compatible} with the action of the operators, thus providing the {\it true 
quantization}.

In fact, from the group law $g''=g'*g$ of any group $\tilde{G}$, we can read 
two different actions:
\begin{eqnarray}
g''&=&g'*g\equiv L_{g'}g\nonumber\\
&&\nonumber\\
g''&=&g'*g\equiv R_{g}g'\nonumber
\end{eqnarray}

\noindent The two actions commute and so do the generators $\tilde{X}^R_a$  and 
$\tilde{X}^L_b$ of the {\it left} and {\it right} actions respectively, 
i.e.
\[[\tilde{X}^L_a,\;\tilde{X}^R_b]=0\;\;\forall a,b\;.\]

\noindent The generators $\tilde{X}^R_a$ are right-invariant vector fields 
closing a Lie algebra, ${\cal X}^R(\tilde{G})$, isomorphic to the tangent space
to $\tilde{G}$ at the identity, $\tilde{\cal G}$. The same, changing $L\leftrightarrow R$, applies
to $\tilde{X}^L\in{\cal X}^L(\tilde{G})$.

We consider the space of complex functions $\Psi$ on the whole group 
$\tilde{G}$ and restrict them to only $U(1)$-functions, that is, those 
which are homogeneous of degree one on the argument $\zeta\equiv e^{i\phi}\in 
U(1)$; that is,
\[ \tilde{X}^L_\phi\Psi= i \Psi\,, \]

\noindent where $\tilde{X}^L_\phi$ is the (central) generator of the $U(1)$ subgroup. On these 
functions the right-invariant vector fields act as {\it pre-Quantum Operators} by ordinary derivation. 
However, this action is not \textit{irreducible} since there is a set of non-trivial operators 
commuting with this representation. 
In fact, all the left-invariant vector fields do commute with the right-invariant ones, i.e. the (pre-quantum) operators. According to Schur's Lemma
those operators must be ``trivialized'' in order to turn the right-invariant vector fields into true \textit{quantum operators}.

We have seen that the action of the central generator (which is, in particular, left-invariant) is fixed to be non-zero by the $U(1)$-function condition. Thus, not every left-invariant vector field can be nullified in a compatible way with this condition. That is, if 

\[
   \left[ \tilde X_{a}^L,\tilde X_{b}^L\right]=\tilde{X}^L_\phi,
\]
then $X_{a}^L \Psi=0, \; X_{b}^L \Psi=0$ is not compatible with $\tilde{X}^L_\phi\Psi= i \Psi$. Of course, 
this null condition on $\Psi$ can be imposed by those generators that never produce a central term by
conmutation; they constitute the \textit{Characteristic sub-algebra} $\mathcal G_{\Theta}$. But also half
of the rest of left-generators can be joined to $\mathcal G_{\Theta}$ to constitute a \textit{Polarization} subalgebra ${\cal P}$.

Then, the role of a Polarization is
that of {\it reducing} the representation, which now constitutes a true {\it Quantization}. Therefore we impose that \textit{wave functions} satisfy the Polarization condition:  
\[ \tilde{X}^L_b\Psi=0\,, \ \ \forall \tilde{X}^L_b\in {\cal P}\, .\]
We refer the reader to Ref. \cite{23} for further details.

\subsection{The case of a particle on the sphere}

The procedure presented above is quite simple when applied to the case of a particle moving on the coset space $\mathbb S^2\equiv SU(2)/U(1)$. We begin with a specific realization of the basic group $\tilde G^{(1)}$, which in this case turns out to be a centrally extended Euclidean Group. In it, the fields $X_{\epsilon^i}\,, i=1,2,3$ generate ordinary rotations, parameterized by vectors $\vec\epsilon$ whose direction determines the axis of rotation and its modulus the angle of rotation by $|\vec\epsilon|=2\sin \frac{\chi}{2} $. In the same way, the fields $X_{\theta^i}\,, i=1,2,3$ correspond to the (co-)tangent subgroup, parameterized by $\theta^i$. In terms of these variables the group law is:
 
 \begin{eqnarray*}
    R(\vec\epsilon\,'') &=& R(\vec\epsilon\,') R(\vec\epsilon\,) \\
    \vec \theta\,'' &=& R^{-1}(\vec\epsilon\,) \vec \theta\,' +  \vec \theta\\
    \zeta'' &=& \zeta' \zeta \exp \left\lbrace 
    i\frac{m r}{\hbar} \vec\lambda \cdot \left( \vec \theta\,''-\vec \theta\,'-\vec \theta\right)\right\rbrace =
    \zeta' \zeta \exp \left\lbrace i\frac{m r}{\hbar} \vec\lambda \cdot \left(R^{-1}(\vec\epsilon) \vec \theta\,'
    - \vec \theta\,'\right)\right\rbrace \\
    \phantom{\zeta''} & \phantom{=} & ( R(\vec\epsilon\,)^i_j \equiv
                                      (1-\frac{{\vec\epsilon}^{\;2}}{2})\delta^i_j-
                                      \sqrt{1-\frac{{\vec\epsilon}^{\;2}}{4}} \eta^i_{\cdot jk}\epsilon^k +
                                      \frac{1}{2} \epsilon^i \epsilon_j \;
                                      ),
 \end{eqnarray*}

\noindent where $\vec\lambda$ is an arbitrary, constant vector in the (co-)algebra with modulus $r$, and $\zeta\equiv e^{i\phi}$ is the quantum mechanical phase. The $\hbar$ constant has been introduced to keep the exponent dimensionless.

We can immediately calculate the corresponding infinitesimal generators of the left action, which are the  right-invariant vector fields:

\begin{eqnarray*}
    \tilde X^{R}_{\vec\epsilon} &=& X^{R\;(SU(2))}_{\vec\epsilon} \\
    \tilde X^{R}_{\vec \theta}       &=& R^{-1} \frac{\partial}{\partial \vec \theta} 
                                    + \frac{m r}{\hbar}(R \vec\lambda - \vec\lambda)^t
                                     \Xi\\
    \tilde X^{R}_{\phi}        &=& \hbox{Re}\left(i\zeta 
                                    \frac{\partial}{\partial\zeta}\right)
                                    \equiv \Xi.
\end{eqnarray*}

\noindent Needless to say that this set of generators closes (a particular case of) the Lie algebra that we found in the previous section:

\begin{eqnarray*}
\left[\tilde X^{R}_{\epsilon^{i}}, \tilde X^{R}_{\epsilon^{j}}\right] 
&=& 
- \eta_{ij\cdot}^{\phantom{ij}k} \tilde X^{R}_{\epsilon^{k}}
\\
\left[\tilde X^{R}_{\epsilon^{i}}, \tilde X^{R}_{\theta^{j}}\right] 
&=&
- \eta_{ij\cdot}^{\phantom{ij}k} \left( \tilde X^{R}_{\theta^{k}} +\frac{m r}{\hbar} \lambda_{k} \Xi \right) 
\\
\left[\tilde X^{R}_{\theta^{i}}, \tilde X^{R}_{\theta^{j}}\right]
&=& 
0\;.
\end{eqnarray*}

We can compute as well the infinitesimal generators of the right action, which are the  left-invariant vector fields:

\begin{eqnarray*}
    \tilde X^{L}_{\vec\epsilon} &=& X^{L\;(SU(2))}_{\vec\epsilon} - 
                                  \vec \theta \wedge \frac{\partial}{\partial \vec \theta}
                                  -\frac{m r}{\hbar}\vec \theta \wedge \vec \lambda \;\Xi \\
    \tilde X^{L}_{\vec \theta}       &=& \frac{\partial}{\partial \vec \theta} \\
    \tilde X^{L}_{\phi}        &=& \hbox{Re}\left(i\zeta 
                                   \frac{\partial}{\partial \zeta}\right)\equiv \Xi,
\end{eqnarray*}

\noindent closing the same Lie algebra but with opposite structure constants.

The characteristic module, i.e., the sub-algebra generated by those vector fields which do not produce a central term under conmutation, and therefore without dynamical content, is generated by two fields:

\[
   \mathcal G_{\Theta}= \langle \vec\lambda \cdot \tilde X^{L}_{\vec\epsilon},\, 
   \vec\lambda \cdot \tilde X^{L}_{\vec \theta}\rangle,
\]

A polarization sub-algebra is given by the characteristic module together with half of the conjugated pairs: 

\[
   \mathcal P = \langle \vec\lambda \cdot \tilde X^{L}_{\vec\epsilon},\, \tilde X^{L}_{\vec \theta}\rangle.
\]

This must not be interpreted as constraint conditions, since they preserve the action of the right-invariant vector fields, i.e., those which will be the physical operators. An alternative treatment, making explicit use of constraints can be seen, for instance, in \cite{coherent}.

An irreducible representation of the group is given by the action of the right-invariant vector fields of the group on the complex functions valued over the group manifold, provided that these functions are polarized and satisfy the condition of U(1)-function:
 
 \[
    \mathcal P \Psi = 0, \qquad \Xi \Psi = i \Psi.
 \]

It can be easily checked that such functions (now true \textit{wave functions}) are of the form

\[
   \Psi \left(R(\vec\epsilon\,)\vec\lambda,\vec \theta, \zeta\right)=\zeta 
    \Phi \left(R(\vec\epsilon\,)\vec\lambda\right),
\]
\noindent where $\Phi$ is an arbitrary function, or equivalently,  in terms of the variable $\vec S \equiv R(\vec\epsilon\,)\vec\lambda$ (the positions on the surface of a sphere of radius $r$):

\[
   \Psi (\vec S,\vec \theta, \zeta)=\zeta \Phi (\vec S).
\]

The explicit action of the right-invariant vector fields on these wave functions is computed to give:

\begin{eqnarray*} 
   \tilde X^{R}_{\vec\epsilon} \Psi &=& \zeta 
           \left(\vec S \wedge \vec \nabla_{S} \Phi \right)= \vec S \wedge \vec \nabla_{S} \Psi  \\
   \tilde X^{R}_{\vec \theta} \Psi       &=& i\frac{m r}{\hbar} \left( \vec S - \vec \lambda \right) \Psi.
\end{eqnarray*}

It is possible to redefine the vector fields to obtain the actual quantum operators, acting only on the arbitrary part of the wave functions.
Thus, we end up with the explicit representation over the wave-functions $\Phi$ depending only in the variable $\vec S = R\vec\lambda$,

\begin{eqnarray*}
   \hat{\vec L} \Phi (\vec S) &\equiv& i \hbar \vec S \wedge \vec \nabla \Phi(\vec S)\\
   \hat{\vec S} \Phi (\vec S) &\equiv& \vec S \Phi(\vec S).
\end{eqnarray*}

\noindent It becomes evident at this point that the domain of the wave functions have been naturally selected without imposing any constraint condition as such.

As was previously emphasized, the $\hat{\vec L}$ operators now play the role of ``generators of translations'' on the surface of the sphere, $\hat{\vec S}$ playing that of a ``position operator'' on the sphere surface\footnote{Note that, in this representation of the Euclidean Group, the relation $\hat{\vec S}\cdot \hat{\vec L}=0$ is fulfilled. However, the Euclidean Group admits another family of central extensions, where $\hat{\vec S}\cdot \hat{\vec L}\neq 0$, and the corresponding representations can realize a magnetic monopole in the center of the sphere. They will be studied elsewhere.}. 

Finally, to obtain a Hamiltonian, we proceed as in the free particle: it is defined as the generator of ``translations'' squared, so that

\[
   \hat H = \frac{\left.\hat{\vec L}\right.^{2}}{2 m} = \frac{\hbar^{2}}{2 m} \nabla^{2}.
\]

There is no ambiguity in this expression, since it corresponds to the squared action of $\hat{\vec L}$ as a ``basic operator''. $\hat{H}$ thus  provides the energy spectrum

\[
   E_l = \frac{\hbar^{2}}{2 m} l (l+1)
\]

\noindent with no extra terms. Note that $\hat H$ coincides with the Casimir
of $SU(2)$ restricted to $\mathbb{S}^2$. In the general case, the energy operator 
will be the quadratic Casimir of $G$ restricted to $G/G_\lambda$, and this turns out
to be the Laplace-Beltrami operator on  $G/G_\lambda$.

\section{Conclusions}

As we have shown with the particular example of the free particle in the sphere $\mathbb S^2$, the quantization of a non-linear system can be achieved if the complete (quantum) symmetry of the system is identified. As a remarkable advantage over other quantization methods, the followed one, a Group Approach to Quantization, is coordinate-free, in such a way that the quantum operators are automatically selected. In particular, this implies that there is no need to impose any constraint. This is specially well suited for non-linear systems, where normal-ordering problems arise in canonical quantization.

We would like to point out that a similar strategy can be explicitly adopted in dealing with particles moving on coadjoint orbits of semi-simple groups, non-necessarily spheres, as shown in our general scheme.

\subsection*{Acknowledgments}

Work partially supported by the Spanish MCYT, Junta de Andaluc\'\i a and Fundaci\'on
S\'eneca under projects FIS2005-05736-C03-01, P06-FQM-01951 and 03100/PI/05. F. F. 
L\'opez Ruiz would like to thank the C.S.I.C. for an I3P grant.

\end{document}